\documentstyle[preprint,prb,aps]{revtex} 
\tightenlines 
\newcommand{\be}{\begin{equation}} 
\newcommand{\ee}{\end{equation}} 
\newcommand{\bea}{\begin{eqnarray}} 
\newcommand{\eea}{\end{eqnarray}}

\begin{document}
\draft

\title{
Random Mass Dirac Fermions in Doped Spin--Peierls and
Spin--Ladder systems: One--Particle Properties and Boundary Effects
}
\author{
M. Steiner$^{(a)}$, M. Fabrizio$^{(b)}$, and
Alexander O. Gogolin$^{(a)}$
}
\address{
$^{(a)}$Imperial College,
Department of Mathematics, 180 Queen's Gate, London SW7 2BZ, U.K.\\
$^{(b)}$International School for Advanced Studies 
Via Beirut 4, 34014 Trieste, Italy \\
and Istituto Nazionale della Fisica della Materia INFM, Italy
}
\date{\today}
\maketitle
\begin{abstract}
Quasi--one--dimensional spin--Peierls and spin--ladder systems 
are characterized by a gap in the spin--excitation spectrum, which
can be modeled at low energies by
that of Dirac fermions with a mass.
In the presence of disorder
these systems can still  
be described by a Dirac fermion model, but with a random mass.
Some peculiar properties,
like the Dyson singularity in the density of states,
are well known and attributed to creation of
low--energy states due to the disorder.
We take one step further and
study single--particle correlations 
by means of Berezinskii's diagram technique.
We find that, at low energy $\epsilon$, 
the single--particle Green function 
decays in real space like
$G(x,\epsilon) \propto (1/x)^{3/2}$.
It follows that at these energies the correlations in
the disordered system are strong -- even stronger than
in the pure system without the gap.
Additionally, we study the effects of boundaries
on the local density of states.
We find that the latter is logarithmically
(in the energy) enhanced close
to the boundary. 
This enhancement decays into the bulk as 
$1/\sqrt{x}$ and the density of states saturates to its
bulk value on the scale 
$L_\epsilon \propto \ln^2 (1/\epsilon)$.
This scale is different from the Thouless localization
length $\lambda_\epsilon\propto\ln (1/\epsilon)$. 
We also discuss some implications of these results 
for the spin systems and their relation to the
investigations based on real--space renormalization
group approach. 
\end{abstract}
\narrowtext

\section{Introduction}

One--dimensional quantum electron (and spin) systems have
attracted considerable attention of theorists over the decades.
The interest to these systems have been particularly boosted
in recent years by the rapid development of experimental
techniques. 
The latter include 
the discovery of various non--organic quasi--one--dimensional
compounds. 
In particular, 
the materials we shall be concerned with in this paper are
the recently discovered $GeCuO_3$ 
spin--Peierls compounds \cite{SP} and the spin--ladder
compounds $(VO)_2P_2O_7$ and $SrCu_2O_3$ \cite{Rice}.
The modern experimental techniques allow measurements on these 
inorganic compounds,
which were either impossible or inaccurate with the 
organic spin--Peierls materials in the past. 
These measurements not only can be now performed on single
crystals, but also involve controlled doping by impurities.
The latter possibility brings up the old issue 
of the disorder effects in one--dimensional systems:
a fascinating subject which was
considered in the past as being somewhat academic.

Among various one-dimensional disordered systems, 
random exchange spin chains have been 
studied, using the so-called real--space renormalization group, 
with very intriguing results \cite{Ma,Fisher,Bhatt}. 
In particular, it has been shown\cite{Fisher}  that 
the typical behavior of
the correlation functions may be quite different from the
average behavior, 
which is more relevant from the experimental point
of view. 
For instance, while the typical correlation functions usually 
decay exponentially, 
the average ones can be power law decaying. 
Moreover, 
even when the average correlation functions decay exponentially, 
the correlation length is different (bigger)
from that of the 
typical correlations\cite{Bhatt}. 
This feature is a consequence of 
rare disorder configurations dominating the 
long distance behavior of the 
correlation functions. 
The analogy 
of these spin systems with the effective disordered
fermionic models have not been investigated
in detail. 

In this paper we shall partly fill this gap, 
and investigate more deeply the fermionic models.
The main reason is that fermions naturally appear in the
description of pure spin--Peierls chains and
especially in the spin ladders. 
Moreover, this alternative 
approach is somewhat complementary to the real space 
renormalization group
analysis, 
thus together they would provide a complete and 
satisfying description of such disordered systems.

Specifically, considering that both the spin--Peierls
and the spin--ladder systems have a spin gap in the
excitation spectrum, it is not surprising that the
effective fermionic model for both systems is
a model of massive Dirac fermions. 
The effect of 
non--magnetic impurities is to randomize the mass
(see also the next Section).
Thus, the ultimate fermionic model, on which we focus 
in what follows, is the one--dimensional random mass Dirac fermion
model. 

A great deal is known about simple self--averaging quantities
for this model, like the total density of states and the 
localization length.

For a Gaussian (white noise) distribution of the mass
variable $m(x)$ the density of states was calculated
by Ovchinnikov and Erikhman \cite{OE}.
Using Fokker--Planck type equations for the probability
distribution of the wave--function phase, they obtained
the divergent density of states \cite{CDM}, 
\be
\rho(\epsilon)\propto
\frac{1}{|\epsilon\ln^3(1/\epsilon)|}\;,
\label{IDOS}
\ee
as the energy approaches the midgap $\epsilon\to 0$.

Physically, the appearance of the
singularity in the density of states can be easily 
understood.
Indeed, for a single kink (sign change) of the mass,
there always is a zero--energy bound state,
localized around the point in space where $m(x)=0$.
For many kinks, there are many such localized states
(also referred to in the literature as zero--modes or solitons).
If they were not overlapping, the density of states
would have a $\delta$--function peak at $\epsilon=0$.
In fact the zero--modes are overlapping. 
Hence the $\delta$--peak broadens, but the
singularity remains -- Eq.(\ref{IDOS}).
[In order to give a qualitative explanation as to
why the broadened $\delta$--peak takes 
exactly the shape (\ref{IDOS}), a more elaborated 
argumentation is needed: see, e.g., Ref.\onlinecite{CDM}.]
It must be noted that
the singularity in the density of states of the form
(\ref{IDOS}) has been discovered by Dyson
back in 1953 \cite{Dyson} for a model of a disordered
harmonic chain.
In the electronic spectrum at the center of the Brillouin
zone such a singularity was identified by
Weissman and Cohan \cite{WC} for the case
of an off--diagonal disorder (random hopping model).
The latter model is in fact directly related to 
the random mass Dirac problem.

According to the Thouless relation \cite{Th},
the localization length corresponding to the
density of states (\ref{IDOS}) takes the form
\be
\lambda_\epsilon\sim l\ln(1/\epsilon)\;,
\label{ILOC}
\ee
where $l$ is the mean free path.
The fact that the localization length 
diverges in the middle of the
gap makes the random mass Dirac problem different 
from the usual one--dimensional localization
problem.
For the latter, the localization length essentially 
coincides with the mean free path $l$, so that
there is only one length scale in the problem.
Consequently, it makes sense to consider only the
distances $x \gg l$ where the correlation functions 
decay exponentially \cite{Ber}.
On the contrary, for the random mass Dirac fermions, various
length scales come into play: 
the mean free path $l$,
the localization length $\lambda_\epsilon$, and
what we shall call the correlation length $L_\epsilon$.
We are going to argue that 
\[
L_\epsilon \sim l \ln^2(1/\epsilon)\;.
\]
(This is in agreement with the renormalization group studies,
mentioned above; see also Section \ref{sec:DC}.)  
The correlation length diverges even faster
than the localization length.
While we do expect that the correlation
functions for the random mass Dirac fermion model
behave in the same way as those for standard 
one--dimensional disordered 
systems at distances $x \gg L_\epsilon$,
there is a different regime $l \ll x \ll L_\epsilon$,
arising at low energies. 
A new physics emerges in this regime, the
understanding of which this paper is intended to 
contribute to.
Specifically, the system seems to exhibit some kind of
criticality -- all the known 
correlation functions are power--law decaying with
universal exponents. 
This behavior is determined by rare fluctuations
when the particle wave--functions have large amplitudes.

As we already mentioned,
the simple self--averaging quantities, like the total
density of states, can be computed by various
methods like the Fokker--Planck type equations
for the probability distributions or the replica 
trick \cite{George}.
All these methods are, however, ill--suited for
determining the correlation functions.
In their recent preprint Shelton and Tsvelik
succeeded to make use of an elegant mapping
of the Dirac problem at zero energy ($\epsilon=0$)
onto the so--called Liouville quantum mechanics \cite{ST}
for calculating the disorder averages of the
products $\psi_0^2(x_1)\psi_0^2(x_2)...\psi_0^2(x_N)$
of the zero--energy wave--function amplitudes.
While this mapping does provide an interesting 
insight into the random mass
Dirac problem (strictly) at zero energy, we are interested to
get more information about the correlation functions,
in particular about the correlations involving the phase
of the wave--functions and about their energy dependence.

On the other hand, a method, which does (in principle) allow
to calculate all the correlation functions, has been
around for almost 25 years. 
The method is the Berezinskii diagram technique originally
invented (back in 1973) to confirm the Mott hypothesis of
$\omega^2\ln \omega$ vanishing conductivity for the
standard one--dimensional localization problem \cite{Ber}.
The Berezinskii method was extended 
by A.A. Gogolin and Mel'nikov (GM) \cite{AAG} to the
case of a half--filled electron band with  a random 
backscattering, the model equivalent to the
random mass Dirac fermions.
Needless to say that GM were able to reproduce the
Dyson singularity in the density of states.
More importantly, they calculated the current--current and
the density--density correlation functions. 
The current--current correlation function 
has no apparent physical meaning for the spin systems.
From the point of view of the latter,
the density--density correlation function is more interesting,
for it turns out to be proportional to the staggered 
susceptibility (at least for spin--Peierls systems, see 
also below). 
GM found that, in the region $l \ll x$,
the contribution to the density--density correlation function at low frequency 
$\omega$ of the state with energy $\epsilon$ such that $x\ll L_\epsilon$ 
decays as
\be
\langle \psi^\dagger_{\epsilon+\omega}(x)
\psi^{\phantom{\dagger}}_\epsilon(x) 
\psi^\dagger_{\epsilon}(0)
\psi^{\phantom{\dagger}}_{\epsilon+\omega}(0) 
\rangle \propto \left(
\frac{1}{x}\right)^{3/2}.
\label{IDDC}
\ee
(This asymptotics was reproduced by Shelton and Tsvelik \cite{ST}
for $\epsilon=\omega=0$). 

In this paper we further study the random mass Dirac problem
following the line pioneered by GM. 
After defining the model and
discussing its relation to spin systems, we 
briefly introduce the Berezinskii technique 
(we feel that that would be convenient
for the reader since a transparent explanation of
the Berezinskii technique is often neglected in the
literature). 
Then we apply the technique to the calculation of the
single--particle Green function at low energy.
Amusingly,
we find that the Green function decays as
$\left(1/x\right)^{3/2}$ (for $l\ll x \ll L_\epsilon$).
Namely, it is characterized by the same exponent as
the density--density correlation function. 

In the second part of the paper we study a 
somewhat different (though related) question. Namely,
we calculate the local density of states 
in the presence of a boundary. 
Unlike the case of the usual localization
[for which the effects of boundaries were
studied by Al'tshuler and Prigodin (AP) \cite{AP}],
for the random mass Dirac problem
one might expect that the boundaries 
affect already the average local density of states
(and not just its distribution function).
Indeed, it turns out that the Dyson singularity is
enhanced close to the boundary as $\ln (1/\epsilon)$, as
compared to its bulk value.
The local density of states then decays into the
bulk of the sample following the law $1/\sqrt{x}$,
and finally saturates to the bulk value at 
$x\sim L_\epsilon$.
We also confirm
these results by a real--space renormalization group 
analysis.
Some consequences for the thermodynamics of the
spin systems are discussed in the last Section. 

We would like to stress that, despite all the results
mentioned above,
we are just starting to study the problem in this paper.
Ultimately, a comprehensive understanding of the
new physics, arising in the random mass Dirac problem
will only be achieved when
the fluctuations of the correlation functions are
understood. 
The problem is mathematically involved, so that,
unlike the usual case \cite{AP}, even the distribution
function of the simplest quantity, the local density of states,
is unknown at present. We hope to return to these 
questions in future publications \cite{SFG}. 

\section{The Model}

The one--dimensional Dirac Hamiltonian is of the form
\be
H_{{\rm D}} = 
\int dx \{v \Psi^{\dagger} (- i \partial_x) \sigma_z \Psi + 
m(x) \Psi^{\dagger} \sigma_x \Psi \},
\label{MDir}
\ee
where $\sigma_x, \sigma_z$ are the Pauli matrices, and
$\Psi^\dagger (x)= \left[ R^\dagger(x),\; L^\dagger(x) \right]$. 
The operators $R(x)$ [$L(x)$] stand for the chiral 
right (left) moving electron fields.
The mass term $m(x)$ is randomly 
distributed.
Throughout this paper we shall assume $m(x)$ be Gaussian
with vanishing average: $\langle m(x)\rangle =0$.  
Its distribution is therefore determined 
by the correlation function
\be
\langle m(x) m(x^{\prime}) \rangle = \frac{v^2}{l} 
\delta(x - x^{\prime})\;,
\label{Mcor}
\ee
where $l$ is the mean free path
(and $v$ is the velocity).

In practice, the Dirac model (\ref{MDir}) usually arises as 
a continuum version of a lattice model, so that $R$ and $L$
are the chiral components of the continuum limit for a lattice
electron field operator
\be
\psi_n\rightarrow {\rm e}^{ik_{F}x}R(x)+
{\rm e}^{-ik_{F}x}L(x),\;\;\;
{\rm for}\;\;\;x=na_0\;.
\label{MCL}
\ee
Here $a_0$ is the lattice spacing and 
the condition of being at the half--filling for
the lattice model means that $4k_Fa_0=2\pi$.
From (\ref{MCL}) it is clear that (\ref{MDir}) is just
another way to write down the random backscattering 
electron model at half--filling. 

Notice that there is no forward scattering in the
problem. 
This is rigorously true for a 
simple tight-binding model at half-filling 
with random nearest-neighbor hopping integrals. 
This model is particle-hole symmetric
\[
\psi^{\phantom{\dagger}}_n \to (-1)^n \psi^\dagger_n,
\]
which prevents any generation of forward scattering processes. 
Hence this model in the continuum limit would reproduce 
the Hamiltonian (\ref{MDir}). 
 
The Hamiltonian (\ref{MDir}) has been shown to also describe
the low-energy limit of 
one--dimensional spin systems with 
a spin gap 
in the excitation spectrum (see Refs\onlinecite{FM,NYG}).
For the convenience of the reader we shall present
the qualitative side of the argument.

\begin{itemize}
\item{} Spin--Peierls systems. The spin part of the Hamiltonian
for a single spin-1/2 chain interacting with phonons
takes the form 
\be
H_{{\rm SP}}=\sum\limits_n J\vec{S}_n\vec{S}_{n+1}+
\sum\limits_n \Delta_n(-1)^n\vec{S}_n\vec{S}_{n+1}\;,
\label{MSPH}
\ee
where $\vec{S}_n$ are spin--1/2 operators, $J$ is the exchange
coupling constant, and $\Delta_n$ measures the strength of
the dimerization caused by the interaction with the phonons.
Below the spin-Peierls transition temperature, one usually
assumes $\Delta_n \rightarrow \langle \Delta_n \rangle=\Delta$
(the mean--field approach is appropriate for the phonons
normally have a three--dimensional dispersion relation).
The passage to the model (\ref{MDir}) is as follows.
The XY--version of the model (\ref{MSPH}) can be mapped 
onto a fermionic model in a standard way by using the
Jordan--Wigner transformation.
Taking the continuum limit leads to (\ref{MDir}) with
the parameter $\Delta$ determining the mass $m_0$. 
The $J_z$ coupling corresponds to a four--fermion interaction
term, but it is irrelevant because of the spin gap.
The main effect of the doping is to randomize the mass term $m(x)$. 
As argued in Ref.\onlinecite{FM}, a single impurity effectively
introduces a domain wall into the system -- a kink
separating phases with different signs of the dimerization order
parameter. 
The mass variable, proportional to the
dimerization strength, will thus also have a kink
at the domain wall position $x_0$, changing
from $m(x)=m_0$ for $x<x_0$ to $m(x)=-m_0$ for $x>x_0$.
\item{} Spin--ladder systems. The Hamiltonian of two
coupled spin--1/2 chains is
\be
H_{{\rm SL}}=\sum\limits_{n,j=1,2}J\vec{S}_{n,j}\vec{S}_{n+1,j}+
\sum\limits_n J_\perp\vec{S}_{n,1}\vec{S}_{n,2}\;,
\label{MSLH}
\ee
where the coupling between the chains, $J_\perp$,
is assumed to be small (this it not qualitatively
restrictive but allows one to construct a consistent theory).
In the continuum limit description,
the spin density operators on each chain have slow--varying
components as well as fast (staggered) components:
\be
\vec{S}_{n,j}\rightarrow \vec{S}(x)+(-1)^n\vec{n}_j(x)\;.
\label{MSop}
\ee
The staggered correlations are stronger than the 
uniform correlations, so the most relevant interaction between
the chains is 
\be
J_\perp a_0^{-1} \int dx \, \vec{n}_1 (x) \cdot \vec{n}_2 (x)
\label{Mnn}
\ee
(other couplings are, in fact, marginal).
The scaling dimension of the coupling (\ref{Mnn}) is equal to
unity. 
Therefore, as it was realized by
Shelton, Nersesyan, and Tsvelik \cite{SNT},
this coupling can be conveniently re--fermionized.
The result is the Dirac model (\ref{MDir}) with the mass
parameter proportional to coupling $J_\perp$ \cite{Mremark1}.
Let us now discuss what the effect of doping would be.
Consider, for example, the situation in 
$La_1CuO_{2.5}$ spin--ladder system
doped by $Sr$ (which substitutes $La$) \cite{LDop}.
For low $Sr$ concentrations ($x<0.02$),
the holes (carrying spin-1/2's) are localized on the Oxygen atoms,
i.e. in between neighboring sites of the $Cu$ magnetic lattice.
As a result the magnetic sites from the right of such an
impurity are effectively re--numbered (shifted by one).
Consequently, the staggered magnetization changes sign
causing a sign change of the coupling (\ref{Mnn}) and
of the mass term of the corresponding Dirac problem \cite{Mremark2}.
\end{itemize}

Thus, the above considerations of the doping effects lead,
for both the spin--Peierls and the spin--ladder systems, to
the random mass Dirac model. 
Specifically, the mass turns out to be distributed as 
a random `telegraph signal'
\be
m(x)=m_0 \prod\limits_i {\rm sgn} (x-x_i)\;,
\ee
$x_i$ being randomly distributed.
At low energies (namely at energies well below the gap: 
$\epsilon \ll m_0$),
the model is still equivalent to the one with a Gaussian 
distribution of the mass \cite{CDM}.
Since we are mainly interested in the asymptotic behavior of the
correlation functions (i.e., in the parameter
region where the correlation 
functions are expected to be universal: $\epsilon\ll m_0$ while 
$l\ll x\ll L_\epsilon$),
we can safely assume the Gaussian distribution.
Then the correlation functions can be calculated
by means of the
Berezinskii diagram
technique \cite{Ber} introduced in the following Section.

\section{Berezinskii diagram technique}

The Berezinskii technique 
is based on the perturbation theory for the one--particle
Schr\"{o}dinger equation (rather than on the many--body
perturbation theory).
Consider a translationally invariant one--particle
Hamiltonian $\hat{H}_0$ perturbed by a random potential
(random mass) term $\hat{m}$.
The bare retarded Green function in the energy representation
is given by 
\be
\hat{G}^{0+}=\frac{1}{\epsilon - \hat{H}_0+i\delta}\;.
\label{Ber:Gnot}
\ee
The exact Green function for a given realization
of the disorder is defined as
\be
\hat{G}^{+}=\frac{1}{\epsilon - \hat{H}_0 - \hat{m}+i\delta}\;.
\label{Ber:G}
\ee
This Green function satisfies the equation
\be
\hat{G}^{+}=\hat{G}^{0+}+\hat{G}^{0+}\hat{m}\hat{G}^{+}\;.
\label{Ber:GGnot}
\ee
Iterating (\ref{Ber:GGnot}) one obtains a
standard perturbative expansion for the Green function.
A typical term of the perturbative series is of the form
\be
\hat{G}^{0+}\hat{m}\hat{G}^{0+}\hat{m}\hat{G}^{0+}...
\hat{G}^{0+}\hat{m}\hat{G}^{0+}\;.
\label{Ber:gt}
\ee

The next step is to take the disorder average. Clearly
only the $2n$-th order terms survive 
(i.e., the terms containing
an even number of impurity potentials).
The disorder average 
$\langle m(x_1)m(x_2)...m(x_{2n})\rangle $
is then given by a sum of products of all 
possible pair correlators (\ref{Mcor}). 
(An additional matrix structure of the
mass term simply ensures that the left-- and right movers 
are interchanged in every scattering event.)
For example, at second order there is just one term
\be
\int d x_1 G^{0+}(x',x_1)G^{0+}(x_1,x_1)G^{0+}(x_1,x)
\label{Ber:sec}
\ee
(The usage of the coordinate rather than the momentum representation 
is vital, as it will become clear shortly.)
At fourth order there are two terms and two integration
variables, etc..

Let us now represent these contributions to the Green function
graphically. 
The bare electron Green functions are represented
by solid lines.
The disorder potential correlators by
`interaction' (wavy) lines, which shall always be
vertically aligned. 
The term (\ref{Ber:sec}) for $x'<x_1<x$ is then
given by the graph Fig.\ref{berfigver}(a).
This graph is also an example of an
elementary `interaction' vertex.
Other contributions are represented by more 
complicated diagrams
where several vertices Fig.\ref{berfigver}(a) and (or) other
vertices are joined.
An elementary consideration of the scattering processes
described by the model (\ref{MDir}) shows that 
Fig.\ref{berfigver}(a-d)
exhausts all the possible vertices. 
Using the many--body terminology, 
there are no forward scattering vertices [simply because there
is no scattering within the same chiral branch in (\ref{MDir})].
The vertices Fig.\ref{berfigver}(c,d) are referred to in the literature as
umklapp scattering vertices.
It is very important that, by construction, the diagrams are
$x$-ordered: $x_1<...x_i<x'<x_{i+1}<...
<x_j<x<x_{j+1}<...<x_{2n}$ \cite{Berremark1}.

As we have already stressed, each scattering event 
interchanges the right-- and the left--moving electrons.
The bare Green functions for them are given by
\be
 G^{0+}_R(x', x; \epsilon) = 
 - \frac{i}{v} \theta(x'-x)
e^{i \frac{\epsilon}{v} (x - x^{\prime})}\;,
\;\;\;
G^{0+}_L(x', x; \epsilon) = 
 - \frac{i}{v} \theta(x'-x)e^{i \frac{\epsilon}{v} (x' - x)}\;,
\label{BerGnotRL}
\ee
where $\theta(x)$ is the step function.
Alternatively, one can use the bare Green function of the combined 
electron field (\ref{MCL})
\be
G^{0+}(x', x; \epsilon) = {\rm e}^{ik_F(x-x')}
G^{0+}_R(x', x;\epsilon) + {\rm e}^{-ik_F(x-x')}
G^{0+}_L(x',x;\epsilon) = 
 - \frac{i}{v} e^{i (k_F+\frac{\epsilon}{v}) |x - x^{\prime}|}\;.
\label{BerGnotbis}
\ee

The next important observation is that,
since all the diagrams are space--ordered,
the Green functions (\ref{BerGnotRL},\ref{BerGnotbis})
are actually factorized in each given graph.
This enables one to split the unperturbed Green function 
$G^{0+}(x_i, x_j;\epsilon) =  - \frac{i}{v} 
\exp\left[i (k_F+\frac{\epsilon}{v}) (x_j - x_i) \right]$ 
(for $x_j > x_i$)
into two factors 
$ \exp\left[-i(k_F+\frac{\epsilon}{v}) x_i\right]$ and  
$\exp\left[+i (k_F+\frac{E}{v}) x_j\right]$, 
and to assign the coordinate 
dependence only to the vertices and to the end points, 
not to the lines. 
As a result, the solid lines just tell us how the vertices
are joined: the analytic expression for any graph 
is given by a product of the vertex (and end point) factors.
Note that the factors corresponding to the vertices
Fig.\ref{berfigver} (a,b), $-1/(2l)$ \cite{Berremark2}
and $-1/l$ respectively,
are $x$--independent. 
The factors $(-1/l)
\exp\left[\pm i(\epsilon/v)x\right]$ corresponding
to the vertices Fig.\ref{berfigver} (c,d) also weakly
depend on $x$ (as $\epsilon \to 0$). 
The reason is that the Fermi wave--vector $k_F$
corresponds to half--filling: ${\rm e}^{i4k_F x}=1$.
As it was realized by GM \cite{AAG}, these 
(umklapp) vertices make the problem different from
the standard localization problem. 
(At large energies, i.e. well away from half--filling, 
these vertices 
strongly oscillate and give a negligible contribution after 
integration over the spatial coordinates is performed \cite{Ber}.)

So far we have merely reformulated the problem.
However, the cardinal idea of the Berezinskii methods is
to provide a way of summing all the diagrams.
The procedure is based on an ingenious classification of
diagrams \cite{Ber}.
Since each diagram is characterized by a fixed number of vertices 
in a definite spatial order, 
one can split it into a product of three integrals over the vertex 
contributions: the 
contributions from the region to the left of $x'$, 
those made in the region to the right of $x$, 
and those made in the central region (between $x'$ and $x$).
These integrals, which are referred to in the
literature as `Berezinskii blocks', 
are then classified by the number of lines  
passing through their boundaries. 
For instance, $\tilde{R}_m(x')$ is the sum of all 
contributions to the left of $x'$ with $4m$ lines, and, equivalently,  
$R_m(x)$ defines the sum of all contributions to the right of $x$.
By joining interaction vertices to Berezinskii's block
of a given order, one finds Berezinskii's blocks of higher
(and lower) order. 
This allows for recurrence relations to be derived.
Instead of trying to give here a general description of
the procedure, we shall illustrate the method at work
while calculating the single--particle Green function
in the next Section.

\section{Single--particle Green function}
\label{sec:GF}

By multiplying the iteration series (\ref{Ber:gt})
with different energies (and spatial coordinates)
and applying the Berezinskii method one can, in principle,
calculate all the correlation functions. 
So, the density--density (and current--current)
correlation functions, which are two--particle
correlation functions, were found by GM \cite{AAG}
(see Section \ref{sec:DC} for discussion).
In this paper we concentrate on the
single--particle properties of the random mass Dirac problem.
We thus apply the Berezinskii technique for calculating
the single particle Green function $G^+(x',x;\epsilon)$.

\subsection{Basic equations}

We start classifying the graphs contributing to the
single--particle Green function by considering all
possible end point configurations. 
Since the electron line can be drawn either to the left
or to the right of each of the two end points $x',x$,
there are four possible configurations, as 
shown in Fig.\ref{gffigend}. 
In going along a diagram, the number of the electron lines 
passing through the current vertical cross section of the 
digram may be changed (by some amount $\Delta g$)
when one encounters an interaction vertex. 
So, the vertices Fig.\ref{berfigver}(a,b) do not change
the number of lines, $\Delta g=0$. 
The vertices Fig.\ref{berfigver}(c,d)
do change the number of lines by $\Delta g=\pm 4$.
An elementary analysis shows that the end point 
configurations Fig.\ref{gffigend}(c,d) would require
$\Delta g=\pm 2$. 
This is impossible with the set of elementary vertices 
at our disposal. 
So, the configurations Fig.\ref{gffigend}(c,d) 
do not contribute to the Green function.

We are therefore left with the two end point configurations 
Fig.\ref{gffigend}(a,b). 
The structure of the Berezinskii
blocks corresponding to these configurations is shown in 
Fig.\ref{gffigbl}(a,b).
The $\tilde{R}$ and $R$ blocks can only have 
multiple of $4$ number of the electron lines.
They are therefore identical for both configurations
Fig.\ref{gffigbl}(a,b).
On the other hand,
the central blocks $Z^{(a)}$ and $Z^{(b)}$ differ.
Indeed, given the left--hand--side block $\tilde{R}_{m'}$
has $4m'$ lines and the right--hand--side block $R_m$
has $4m$ lines,
the central block $Z^a$ must have $4m'+1$ ($4m+1$) lines
on the left (right) while the central block $Z^b$
must have $4m'-1$ ($4m-1$) lines on the left (right).
[Thus, $m\geq 0$ for 
$a$--configuration while $m\geq 1$ for $b$--configuration.]

In terms of the Berezinskii blocks the single--particle
Green function is expressed as follows ($x'<x$)
\bea
G^{+}(x^{\prime}, x) = &-& \frac{i}{v} \left\{
\sum\limits_{m, m^{\prime}=0}^{\infty} 
\tilde{R}_{m^{\prime}}(x^{\prime}) 
Z^{(a)}_{m^{\prime}, m}(x^{\prime},x) R_{m}(x)\right.
\nonumber\\
&+& \left.\sum_{m, m^{\prime}=1}^{\infty} 
\tilde{R}_{m^{\prime}}(x^{\prime}) 
Z^{(b)}_{m^{\prime}, m}(x^{\prime},x) 
R_{m}(x)
\right\}\;.
\label{GFGblocks}
\eea

Consider now the Green function for coinciding points $x'=x-0$.
When the end points merge,
the region of the spatial integration in the central blocks
collapses so that only the trivial diagrams for $Z^{(a,b)}$
survive (those which do not include any interaction vertices).
Hence the boundary conditions
\be
Z^{(a,b)}_{m^{\prime}, m}(x-0,x) = \delta_{m^{\prime}, m}
\mbox{\, for \,} m \ge 0 (1)\;.
\label{GF:Zbound}
\ee
For an infinite sample, the boundary conditions to be imposed
on $R_m$'s are
\be
\tilde{R}_0(x)=R_0(x)=1\;.
\label{GF:Rbound}
\ee
Thus, the equal--point Green function is given by
\be
G^{+}(x-0, x) = - \frac{i}{v} \left\{1+
2\sum_{m=1}^{\infty} 
\tilde{R}_{m}(x) 
R_{m}(x)
\right\}\;.
\label{GF:Gegpoints}
\ee
The local density of states is defined as
\be
\rho(x,\epsilon)=-\frac{1}{\pi}{\rm Im}G^+(x-0,x;\epsilon)\;.
\label{GF:DOSdef}
\ee
In an infinite system the quantities (\ref{GF:Gegpoints})
and (\ref{GF:DOSdef}) do not actually depend on $x$,
for the disorder average effectively restores translation
invariance.
(The situation is different for
systems with boundaries we discuss in Section \ref{sec:BE}.)
Note the appearance of the factor of two in front of the sum
in the equation (\ref{GF:Gegpoints}). 
This factor is due to configurations Fig.\ref{gffigend}
mirror imaging each other when the end points merge, 
as illustrated in Fig.\ref{gffigmir}.

The expression (\ref{GF:DOSdef}) was used by GM for calculating the
density of states. 
They found the following differential recurrence relation for
the Berezinskii blocks $\tilde{R}_m$
\bea
\frac{d\tilde{R}_m(x)}{dx} =& -&
\frac{1}{l} \left\{ 4 m^2 \tilde{R}_m (x)+
m(2m-1) {\rm e}^{-i \frac{4\epsilon}{v} x} 
\tilde{R}_{m-1} (x)\right.\nonumber\\&+ &\left.
m(2m +1) {\rm e}^{- i \frac{4\epsilon}{v} x} \tilde{R}_{m+1}(x) 
\right\} \;.
\label{GF:Rdif}
\eea
The right--hand--side blocks
$R_{m}(x)$ mirror the blocks $\tilde{R}_m(x)$,
so they satisfy (\ref{GF:Rdif}) with $x\to-x$.
An integral equation equivalent to the equation (\ref{GF:Rdif}) 
is obtained by joining the 
elementary vertices to $\tilde{R}_{m}(x)$
in all possible ways,
but avoiding creation of closed electron loops. 
(The loop diagrams are not allowed because we study a
one--particle problem. 
Put another way, they vanish because they involve
a retarded Green function.)
The process is illustrated in Fig.\ref{gffigpr}.
The first term in (\ref{GF:Rdif}) comes from
joining the vertices Fig.\ref{berfigver}(a,b)
\[
4m^2=\frac{1}{2}4m+2m(2m-1)\;.
\]
The second and the third term derive from
joining the vertices Fig.\ref{berfigver}(c) and (d)
respectively. 

Carrying out an analogous calculation (Fig.\ref{gffigpr})
for the central blocks $Z^{(a,b)}$,
we obtained the differential recurrence relations
satisfied by these blocks
(in computing the integer coefficients one must be careful
as to not allowing for the electron loops in the whole
diagram):
\bea
\frac{dZ^{(a)}_{*, m}(x)}{dx}  &= &
i \frac{\epsilon}{v} Z^{(a)}_{*, m}(x) - 
\frac{8m^2+4m+1}{2l}Z^{(a)}_{*, m}(x)\label{GF:Zadif}\\
&-&
\frac{m (2m-1)}{l} {\rm e}^{- i\frac{4\epsilon}{v}x}
Z^{(a)}_{*, m-1}(x)-  
\frac{(m+1)(2m+1)}{l} {\rm e}^{ i\frac{4\epsilon}{v} x} 
Z^{(a)}_{*, m+1}(x)
\nonumber
\eea
and
\bea
\frac{dZ^{(b)}_{*, m}(x)}{dx}  &= &
-i \frac{\epsilon}{v} Z^{(b)}_{*, m} (x)- 
\frac{8m^2-4m+1}{2l}Z^{(b)}_{*, m}(x)\label{GF:Zbdif}\\
&-&
\frac{(m-1) (2m-1)}{l} {\rm e}^{- i\frac{4\epsilon}{v}x}
Z^{(b)}_{*, m-1}(x)-  
\frac{m(2m+1)}{l} {\rm e}^{ i\frac{4\epsilon}{v} x} 
Z^{(b)}_{*, m+1}(x)
\nonumber
\eea
Only the active spatial variable $x$ is shown (the variable
$x'$ is suppressed), $*$ stands for the index $m'$ which
plays a role of a parameter.
Defining the Berezinskii blocks [c.f. (\ref{GFGblocks})],
we have incorporated the end point exponentials
$\exp\left(\pm i\frac{\epsilon}{v}x\right)$ into $Z^{(a,b)}$. 
Hence the first terms on the right--hand--side
of the above equations.
[The absence of ${\rm e}^{\pm ik_F x}$ factors means that we are 
actually calculating $G_R$, the total Green
function can be restored by using the
formula (\ref{BerGnotbis}).]

As a next step it is instructive to pass to the momentum
representation for the Green function.
Before doing so we notice that the spatial dependence
of the side blocks can easily be separated \cite{AAG}. 
Indeed, the substitution 
\be
\tilde{R}_m(x) = (-1)^m {\rm e}^{-i \frac{4\epsilon}{v} mx} R_m  
\label{GF:Rsub}
\ee
reduces the equations (\ref{GF:Rdif}) and (\ref{GF:Rbound})
to the algebraic recurrence relation
\be
i s R_m = 4m R_m -(2m-1) R_{m-1} -(2m+1) R_{m+1}  
\label{GF:Rrec}
\ee 
for the $x$--independent quantities $R_m$ obeying
the boundary condition
\be
R_0 = 1\;.
\label{GF:Rboundbis}
\ee
We have introduced the dimensionless energy variable
\be
s=\frac{4\epsilon l}{v}\;.
\label{GF:s}
\ee
It will also be of convenience to pass to the
dimensionless spatial coordinate
\be
y=\frac{x}{l}\;.
\label{GF:y}
\ee

Now we define the objects
\be
Q_m^{(a,b)}(\kappa, s) = l(-1)^m
\sum\limits_{m^{\prime} = 0 (1)}^{\infty}
(-1)^{m'}\int\limits_{y^{\prime}}^{\infty} dy
{\rm e}^{-i\kappa(y - y^{\prime})
+is(my-m'y')}
Z^{(a,b)}_{m^{\prime}, m}(ly^{\prime},ly) 
R_{m^{\prime}}\;.
\label{GF:Q}
\ee 
The sum over $m'$ in the above formula
starts with $0$ for $a$--configurations
and with $1$ for $b$--configurations respectively.

In terms of the objects (\ref{GF:Q}),
the Fourier transformed Green function takes the simple
form 
\bea
G_R(\kappa,s)&=&l\int\limits_{y^{\prime}}^{\infty} dy
{\rm e}^{-i\kappa(y-y')}
G_R(ly',ly;s)=\nonumber\\
&-&\frac{i}{v}
\left[
\sum\limits_{m = 0}^{\infty} 
R_m 
Q_m^{(a)}(\kappa ,s) +
\sum\limits_{m = 1}^{\infty} 
R_m 
Q_m^{(b)}(\kappa, s)\right]\;.
\label{GF:GFourier}
\eea

Finally, applying the operation (\ref{GF:Q})
to Eqs.(\ref{GF:Zadif},\ref{GF:Zbdif}) 
and using the boundary conditions 
(\ref{GF:Zbound}), one can derive algebraic
recurrence relations for  
$Q_m^{(a,b)}(\kappa ,s)$.
We found
\bea
&~&~i s \left(m + \frac{1}{4}\right) Q_m^{(a)} 
- i \kappa Q_m^{(a)} 
- \frac{8m^2+4m+1}{2} Q_m^{(a)}+\nonumber \\ 
&~&~ m (2m-1)Q_{m-1}^{(a)} 
+ (m+1) (2m+1) Q_{m+1}^{(a)} = R_m 
\label{GF:Qarec}
\eea
and
\bea
&~&~i s \left(m - \frac{1}{4}\right) Q_m^{(b)} 
- i \kappa Q_m^{(b)} 
- \frac{8m^2-4m+1}{2} Q_m^{(b)}+\nonumber \\ 
&~&~ (m-1) (2m-1)Q_{m-1}^{(b)} 
+ m(2m+1) Q_{m+1}^{(b)} = R_m \;,
\label{GF:Qbrec}
\eea
where we have suppressed the inactive variables $(\kappa, s)$.

\subsection{Asymptotic solution of the equations}

In order to compute the Green function we need to solve
the recurrence relations 
(\ref{GF:Rrec},\ref{GF:Qarec},\ref{GF:Qbrec})
for the quantities $R_m$ and $Q^{(a,b)}_m$
and use the formula (\ref{GF:GFourier}).

As we already mentioned,
the recurrence relation (\ref{GF:Rrec})
for $R_m$'s was derived by GM.
They also discovered that the
generating function $R(\zeta)=\sum_m R_m\zeta^m$
satisfies a second order differential
equation that happens to be of a hypergeometric
type. 
It can therefore be explicitly solved by, e.g.,
contour integrals method.
The solution found by GM reads \cite{AAG}
\be
R_m = \frac{e^{is/4}}{K_0(-is/4)} 
\int_0^{\infty} \frac{dt}{[t(t+1)]^{1/2}} 
\left(\frac{t}{t+1}\right)^{m} e^{ist/2},
\label{GF:Rm}
\ee
where $K_0$ is the MacDonald function. 
(It can be checked that the 
boundary condition $R_0=1$ is fulfilled.)

The recurrence relations (\ref{GF:Qarec},\ref{GF:Qbrec})
for $Q^{(a,b)}_m$'s are of a more complicated nature. 
So, the second order differential equations satisfied by
the corresponding generating functions $Q^{(a,b)}(\zeta)$ are
more complicated than the hypergeometric equation.
(These equations have four singular points; we have derived
them but we do not feel that they are 
{\it in situ} here.)
Unfortunately, we have not been able to solve these 
equations.

On the other hand, we are mostly interested in the
low--energy regime $s\to 0$.
There one can simplify the recurrence relations 
and change over to differential equations.
Indeed, in the low--energy limit the high
order diagrams matter, so that
we shall restrict ourselves to the parameter region 
\be
s \rightarrow 0, \, m\rightarrow \infty\, 
\mbox{\,\,\,while\,\,\,} \, sm \mbox{\,\,\,finite}.
\label{GF:limit}
\ee

An inspection of the formula (\ref{GF:Rm}), shows that
in the limit (\ref{GF:limit}) 
\be
R(z) = - \frac{2}{\ln(-is)} K_0(z)\;,
\label{GF:Ras}
\ee
where we have introduced the variable
\be
z = \sqrt{-2ims}\;.
\label{GF:z}
\ee
[For large $m$ we can replace $[t/(t+1)]^{m}$ by $\exp(-m/t)$ in
(\ref{GF:Rm}) and the remaining integral, upon
rescaling the integration variable, gives the MacDonald function.]

An alternative way to obtain this approximate solution is to 
pass to the limit of large $m$ in (\ref{GF:Rrec}) \cite{AAGbis}.
Expanding $R_{m\pm 1}$ and neglecting the terms vanishing
in the limit (\ref{GF:limit}), one finds
\[
m^2 \frac{d^2R_m}{dm^2} +m  \frac{dR_m}{dm}+
\frac{i}{2}sm R_m  = 0.
\]
Making the substitution (\ref{GF:z}) one arrives
at the modified Bessel equation
\[
\frac{d^2R}{d^2 z}+\frac{1}{z}\frac{dR}{dz}-R=0\;.
\]
A general solution to this equation is 
$R(z)=AK_0(z)+BI_0(z)$. 
The coefficient $B$ must vanish
for the function $I_0(z)$ diverges at infinity.
The coefficient $A$ is to be determined by the boundary 
condition (\ref{GF:Rboundbis}).
Since the MacDonald function diverges as $z\to 0$,
we must recall that the variable $z$ is discrete. 
There is a $z_{{\rm min}}=\sqrt{-2is}$,
so that $A=1/K_0(z_{{\rm min}})\simeq-2/\ln(-is)$,
in agreement with (\ref{GF:Ras}) within
the leading logarithmic accuracy.
(Within this approximation one must generally neglect 
$C$-numbers as compared to the $log$'s. Yet we are
keeping the imaginary part of the $log$'s. 
Alternatively, the imaginary parts
can be restored at the end of the
calculation by a simple analytic continuation,
i.e. demanding that the Green function be retarded.)

Let us now determine the quantities $Q_m^{(a,b)}$
in the limit (\ref{GF:limit}).
The recurrence relations (\ref{GF:Qarec}) and (\ref{GF:Qbrec})
become identical for large $m$ and reduce to
\be
2 m^2 \frac{d^2Q_m^{(a,b)}}{dm^2} + 4 m \frac{dQ_m^{(a,b)}}{dm}
+\left(\frac{1}{2} - 
i \kappa +ism\right) Q^{(a,b)}_m 
=R_m
\label{GF:Qint}
\ee
Since the summation over $m$ in formula (\ref{GF:GFourier})
is now to be changed to the integration over $z$
[according to $\sum_{m=0(1)}^\infty\rightarrow
(i/s)\int_0^\infty zdz$],
the Green function is determined by the sum $Q=Q^{(a)}+Q^{(b)}$.
The equation for the latter follows from (\ref{GF:z}) and
(\ref{GF:Qint}):
\be
z^2  \frac{d^2Q}{dz^2} + 3 z  \frac{dQ}{dz} 
 -(z^2 + 2i\kappa- 1) Q(z) =4 R(z)\;.
\label{GF:Qz}
\ee
Rewriting the equation (\ref{GF:Qz}) in the form
\be
\frac{d}{dz}\left[z\frac{d(zQ)}{dz}\right]-
\left(z+\frac{2i\kappa}{z}\right)(zQ)=4R(z)\;,
\label{GF:Qzeq}
\ee
we identify it as an inhomogeneous modified Bessel equation for the 
function $z\,Q(z)$.

Equations of this type are usually solved by means of the
Lebedev--Kontorovich transformation \cite{LK}. 
We employ the Lebedev--Kontorovich transformation in the form
\be
\hat{Q}(\tau)=\int\limits_0^\infty dz K_{i\tau}(z)Q(z)\;,
\label{GF:QLK}
\ee
where $K_{i\tau}(z)$ is the MacDonald function of
a purely imaginary index.
The transformation inverse to (\ref{GF:QLK}) is
\be
Q(z)=\frac{2}{\pi^2z}\int\limits_0^\infty 
d\tau \tau \sinh (\pi\tau)K_{i\tau}(z)\hat{Q}(\tau)\;.
\label{GF:QLKinv}
\ee

Applying the integral operator (\ref{GF:QLK})
to the equation (\ref{GF:Qzeq}), re--grouping the
terms and then using the inverse 
transformation (\ref{GF:QLKinv}), 
we find the solution in the form
\be
Q(\kappa,z) = \frac{2}{\pi^2 z} 
\int\limits_0^{\infty} d\tau \, \tau \sinh(\pi\tau)
\frac{K_{i\tau}(z)}{\tau^2 + 2i\kappa} 
\int\limits_0^{\infty} d\xi \, K_{i\tau}(\xi) 4R(\xi)\;,
\label{GF:Qsol}
\ee
where we have restored the dependence of the function $Q$
on the parameter $\kappa$.

At this point we notice that (\ref{GF:Qsol}) is normalized as
\be
\int\limits_{-\infty}^{\infty} \frac{d\kappa}{2 \pi} 
Q(\kappa,z) = 2 R (z)\;.
\label{GF:Qnor}
\ee
According to the definition (\ref{GF:Q}), the
normalization (\ref{GF:Qnor}) corresponds to the
boundary conditions (\ref{GF:Zbound}).

Substituting the asymptotic form (\ref{GF:Ras}) into
(\ref{GF:Qsol}) and looking the $\xi$--integral up in 
Ref.\onlinecite{GR}, we finally obtain
\be
Q(\kappa,z) = - 
\frac{8 }{z \ln(-is)} 
\int\limits_0^{\infty}  d\tau \, \tau
\tanh \left( \frac{\pi\tau}{2} \right) 
\frac{K_{i\tau}(z)}{\tau^2 + 2i\kappa}\;.
\label{GF:Qsolbis}
\ee

\subsection{Calculation of Green's function}

It is now instructive to
Fourier transform the Green function (\ref{GF:GFourier}) back 
to the coordinate space:
\be
G^{+}_R(y, s) = \frac{1}{vs} \int_0^{\infty} dz \, z R(z) Q(y,z)\;.
\label{GF:Gspace}
\ee
The function $Q$ in the space domain is of the form
\be
Q(y,z) = -  
\frac{4}{z \ln(-is)}  
\int\limits_0^{\infty}  d\tau \, \tau
\tanh \left( \frac{\pi\tau}{2} \right) 
K_{i\tau}(z) e^{-\frac{\tau^2}{2} y}\;.
\label{GF:Qspace}
\ee

The $z$--integration is performed in Appendix A.
The final result for the Green function is
\be
G^{+}_R(y,s) = \frac{4}{v s \ln^2(-is)} F(y)\;,
\label{GF:Gfin}
\ee
where the function $F(y)$ is defined by
\be
 F(y)=2\int_0^{\infty} dx \, x \tanh x(1 - \tanh^2 x)  
e^{-\frac{2}{\pi^2} y x^2}.
\label{GF:F}
\ee
The limiting behavior of the function $F(y)$ is
\be
F(y) = \left\{
\begin{array}{lcr}
1 - \frac{1}{2}y+ \mbox{O}(y^2)\;,
&{\rm for}& y\ll 1 \\
2\pi^2  \left( \frac{\pi}{2 y} \right)^{3/2}+ 
\mbox{O}(y^{5/2})\;, &{\rm for}& y\gg 1
\end{array}
\right.
\label{GF:Flim}
\ee
The function $F(y)$ is plotted in Fig.\ref{gffigF}. 
Notice that the correct density of states (Dyson singularity)
is reproduced as
$y\to 0$:
\be
\rho(s)= -\frac{1}{\pi}{\rm Im}G^{+}_R(y\to 0,s)
\simeq \frac{8\pi\rho_0}{s\ln^3(1/s)}\;,
\label{GF:DOS}
\ee
where $\rho_0=1/(2\pi v)$ is the bare density of states
(in the chiral branches of the fermion spectrum).

\section{Boundary effects}
\label{sec:BE}

In this section we study the influence of boundaries
onto single--particle properties of the random mass
Dirac problem. 
This study is largely inspired by the investigation
of the boundary effects for the standard
one--dimensional localization by AP\cite{AP}.
AP succeeded in calculating the whole distribution 
function of the local density of states
(for an infinite system as well as in the presence 
of boundaries). 
Technically, this success was due to a factorization of
the high--order correlation functions into
combinations of known Beresinskii blocks \cite{AP,Ber}.
Unfortunately, for the random mass Dirac problem
such a factorization does not occur.
So, at the time being we do not know the 
distribution functions.
On the other hand, as we shall see shortly,
already the simplest quantity -- the local
density of states -- displays quite an 
interesting boundary behavior for the Dirac problem.

\subsection{Modification of the diagram technique}

We consider a semi--infinite sample ($x>0$),
i.e. we assume an infinite potential wall
to be situated at the origin (the electrons can not
leave the sample).
As a result of the total reflection from the boundary the 
single--particle
eigenfunctions of the Hamiltonian without the disorder 
potential change over
from the plane waves to the standing waves:
\[
\frac{1}{\sqrt{L}} {\rm e}^{ipx}
\rightarrow
\sqrt{\frac{2}{L}}\sin(px)\;.
\]
The bare Green function therefore takes the form
\bea
G^{0+}_{{\rm open}}(x',x;\epsilon) & = & \frac{1}{2}
\left[
G^{0+}(x-x',\epsilon)+
G^{0+}(x'-x,\epsilon)\right. \nonumber\\
&-&\left.
G^{0+}(x+x',\epsilon)-
G^{0+}(-x-x',\epsilon)\right]\nonumber\\&=&
G^{0+}(x-x',\epsilon)-
G^{0+}(x+x',\epsilon)\;.
\label{B:Gnotopen}
\eea
The last equality follows from the fact that 
$G^{0+}(-x)=G^{0+}(x)$ for the translationally
invariant system, Eq.(\ref{BerGnotbis}).

Let us take a closer look at the second term in the 
expression (\ref{B:Gnotopen})
which is due to the presence of the boundary. 
This term can be written as 
\be
-G^{0+}(x+x',\epsilon)=\frac{i}{v}{\rm e}^{i(k_F+\epsilon/v)(x+x')}=
\left\{ -1 \right\}
\left\{ \sqrt{-\frac{i}{v}}{\rm e}^{i(k_F+\epsilon/v)x'} \right\}
\left\{ \sqrt{-\frac{i}{v}}{\rm e}^{i(k_F+\epsilon/v)x} \right\}\;.
\label{B:secterm}
\ee
In terms of the Berezinskii diagrams this term can be 
represented as shown in
Fig.\ref{bfigbv}. 
The second and the third factors 
in curly brackets in the 
expression (\ref{B:secterm}) correspond to the external 
vertices $x'$ and
$x$ respectively, while the $\{-1\}$ factor must be 
attached to the new element of
the diagram technique -- the `boundary vertex' at $x=0$. 
Physically this vertex 
can be interpreted as a total reflection from 
the boundary. (Any phase
factor, like $-1=\exp(i\pi)$, corresponds to a shift of the 
position of the
boundary and can be omitted.)

Thus, in the presence of a totally reflective 
boundary the Berezinskii diagram
technique is modified by adding the 
boundary vertex: Fig.\ref{bfigbv}.

\subsection{Density of states}

The density of states at the point $x$ is given by the general
formula (\ref{GF:DOSdef}).
We must therefore consider the one--loop diagrams: 
the electron line starting at the
point $x$ returns, after undergoing scattering 
processes and reflections from the
boundary, to the same point $x$.
An example of a diagram 
involving the boundary scattering 
is shown in Fig.\ref{bfigex}.
Notice that the boundary vertex changes the
number of electron lines by $\Delta g=2$.
Therefore, in addition to the end point configurations
Fig.\ref{gffigend}(a,b) at work for the infinite system,
the configurations Fig.\ref{gffigend}(c,d) are also allowed.

The diagram Fig.\ref{bfigbv}, in combination with 
the bare Green function, determines the density of 
states in the pure system:
\be
\rho_0(x,\epsilon)=\rho_0\left\{ 1-\cos \left[2 
\left(k_F+\frac{\epsilon}{v}\right)
x\right]\right\} \;.
\label{B:DOSnot}
\ee
The $2k_F$--oscillation of the density of states is
due to the fact that the 
single--electron eigenfunctions are 
standing waves and is not of great interest.
We can get rid of this oscillation by 
adopting the following averaging
procedure 
\be
\rho(x,\epsilon)\rightarrow
\bar{\rho}_\Delta (x,\epsilon)=
\frac{1}{2\Delta}\int\limits_{-\Delta}^{\Delta}d y
\rho(x+y,\epsilon)\;.
\label{B:avp}
\ee
Such averaging over the spatial coordinate has also been discussed 
by AP \cite{AP}.
The parameter $\Delta$ in Eq.(\ref{B:avp}) 
is chosen in such a way that
\be
\frac{1}{k_F}\ll\Delta\ll\min\left(l,\frac{v}{\epsilon}\right)\;.
\label{B:Delta}
\ee
With this choice of $\Delta$ the 
$2k_F$ oscillation in (\ref{B:DOSnot}) averages away
so that
\be
\bar{\rho}_\Delta(x,\epsilon)=\rho_0\left[1+
\mbox{O}\left(\frac{1}{k_F\Delta}\right)\right]\;.
\label{B:DOSnotav}
\ee

It is important to notice that all the 
diagrams with the end point configurations Fig.\ref{gffigend}(c,d),
however complicated, have
an additional rapidly oscillating factor $\exp(\pm 2k_F x)$
as compared to the diagrams with end point
configurations Fig.\ref{gffigend}(a,b).
Upon averaging (\ref{B:avp}), 
the former diagrams acquire
[as (\ref{B:DOSnotav})]
a small factor of the order of $\sim 1/(k_F\Delta)$.
We shall therefore neglect these diagrams. 
(We also observe that the remaining
diagrams do not depend on $\Delta$, 
so we drop this subscript in what follows.)

Defining the Berezinskii blocks in full analogy to the 
translationally invariant case,
we rewrite the expression (\ref{GF:DOSdef}) 
for the density of states in the form
\be
\bar{\rho}(x,\epsilon)/\rho_0=1+ 
2{\rm Re}\sum\limits_{m=1}^\infty \tilde{R}_m(x)R_m(x)\;.
\label{B:DOSbloks}
\ee
[The mirror imaged diagrams 
like Fig.\ref{gffigmir} remain such
even if the boundary vertex is involved (with respect to the
horizontal axes).
Hence the factor of 2 in (\ref{B:DOSbloks}).]

The right--hand--side blocks $R_m(x)$ are not 
influenced by the boundary at all. 
Hence they are identical to those for 
the translationally invariant system (Section \ref{sec:GF}).
As to the left--hand--side blocks $\tilde{R}_m(x)$, 
the presence of the boundary vertex on the left can not
alter the process of constructing this diagrams by 
adding impurity vertices from the right.
Hence $\tilde{R}_m$'s satisfy the same 
differential recurrence relation as they do for the 
infinite system: Eq.(\ref{GF:Rdif}).
What is new in the open boundary case is that a 
boundary condition at $x=0$ comes into play.
Indeed, consider the quantity $\tilde{R}_m(x\to 0)$. 
Tending $x$ to zero means squeezing the
diagrams $\tilde{R}_m(x)$ so that only 
$2m$ boundary vertices (and no impurity vertices)
contribute to $\tilde{R}_m(0)$. 
Hence the boundary condition
\be
\tilde{R}_m(x=0)=1\;\;\;{\rm for}\;\;\;{\rm all}\;\;\; m\;.
\label{B:bc}
\ee

Next we define [c.f. \ref{GF:Rsub}]
\be
\left\{
\begin{array}{l}
R_m(x)=(-1)^m {\rm e}^{i\frac{4\epsilon}{v} mx}R_m\;,\\
\tilde{R}_{m}(x)=(-1)^m{\rm e}^{-i\frac{4\epsilon}{v} mx}L_m(y)\;.
\end{array}
\right.
\label{B:Ldef}
\ee
The Berezinskii blocks $R_m$ are defined in 
Section \ref{sec:GF} [Eq.(\ref{GF:Rm})],
while the blocks $L_m(y)$ satisfy
\be
\left\{
\begin{array}{l}
\displaystyle\frac{d L_m}{d y}
=m\left\{isL_m-\left[
4mL_m-(2m-1)L_{m-1}-(2m+1)L_{m+1}\right]\right\}\;,\\
L_m(y=0)=1\;,
\end{array}
\right.
\label{B:Leq}
\ee
Equation (\ref{B:DOSbloks}) now reads
\be
\frac{\bar{\rho}(y,s)}{\rho_0}=1+ 2{\rm Re}
\sum\limits_{m=1}^\infty L_m(y)R_m\;.
\label{B:DOSblbis}
\ee

Equation (\ref{B:Leq}) 
[in combination with Eq.(\ref{B:DOSblbis}) and Eq.(\ref{GF:Rm})] 
determines, in principle, the exact density of states. 
Unfortunately, 
we failed to obtain the solution to 
the differential recurrence relation (\ref{B:Leq}) in a closed form.
[Upon making the Laplace transform with respect to $y$,
one can reduce (\ref{B:Leq}) to an algebraic
recurrence relation. That is, however,
not any simpler than the relations 
(\ref{GF:Qarec},\ref{GF:Qbrec}) for $Q^{(a,b)}$.]
On the other hand, 
we are primarily interested in the low--energy ($s\ll 1$) 
behavior of the density of states.
So, as in Section \ref{sec:GF},
it is sufficient to consider the limit (\ref{GF:limit}).
For large $m$,
the relation (\ref{B:Leq}) 
reduces to the differential equation
\be
\frac{2}{z}\frac{\partial L}{\partial y}=
\frac{\partial}{\partial z}\left( z 
\frac{\partial L}{\partial z}\right)- zL\;,
\label{B:Lqeq}
\ee
(We recall that $z=\sqrt{-2ims}$.)
Since the operator on the right--hand--side of (\ref{B:Lqeq}) 
is the operator
defining the modified Bessel equation, 
we again (as in Section \ref{sec:GF}) use the 
Lebedev--Kontorovich transformation \cite{LK}.
It is convenient to define
\be
\left\{
\begin{array}{l}
F(y,\tau)=\int\limits_0^\infty\frac{d z}{z}K_{i\tau}(z)L(y,z)\;,\\
L(y,z)=\frac{2}{\pi^2}\int\limits_0^\infty 
d\tau\tau\sinh(\pi\tau)K_{i\tau}(z)F(y,\tau)\;.
\end{array}
\right.
\label{B:LK}
\ee
Applying the transformation (\ref{B:LK}) 
to the equation (\ref{B:Lqeq}) we find that the
function $F(y,\tau)$ satisfies the equation
\be
\frac{\partial F(y,\tau)}{\partial y}=
-\frac{1}{2}\tau^2 F(y,\tau)\;.
\label{B:Feq}
\ee
The solution of this equation is
\be
F(y,\tau)=F_0(\tau){\rm e}^{-y\tau^2/2}\;.
\label{B:Fsol}
\ee
The function $F_0(\tau)=F(0,\tau)$ is 
determined by the boundary condition (\ref{B:Leq})
\be
F_0(\tau)=\int\limits_0^\infty\frac{d z}{q}K_{i\tau}(z)L(0,z)=
\frac{\pi}{2\tau\sinh(\pi\tau/2)}\;.
\label{B:Fnot}
\ee
Combining (\ref{B:Fnot},\ref{B:Fsol}), 
and (\ref{B:LK}) we find the function
$L(y,z)$ in the form
\be
L(y,z)=\frac{2}{\pi}\int_0^\infty d 
\tau \cosh(\pi\tau/2){\rm e}^{-y\tau^2/2}K_{i\tau}(z)\;.
\label{B:Lsol}
\ee
This equation can be re--written in a simpler way as
\be
L(y,z)=\sqrt{\frac{2}{\pi y}}\int\limits_0^\infty d x 
\cos(z\sinh x)\exp\left(-\frac{x^2}{2y}\right)\;.
\label{B:Lsolbis}
\ee
Substituting this expression into the formula (\ref{B:DOSblbis}), 
replacing the sum
over $m$ by the integral over $q$ and 
using the asymptotic expression (\ref{GF:Ras})
for $R_m$'s, we find the density of states in the form
\be
\bar{\rho}(y,s)=\frac{2\pi\rho_0}{s\ln^2(1/s)}f(y)\;.
\label{B:BDOSfin}
\ee
The function $f(y)$ is given by 
\be
f(y)=\sqrt{\frac{2}{\pi y}}\int\limits_{0}^{\infty}
\frac{d x}{\cosh^2x}
(1-x\tanh x)\exp\left(-\frac{x^2}{2y}\right)\;.
\label{B:f}
\ee
It is plotted in Fig.\ref{bfigf}.
The limiting values of the function $f$ are
\be
f(y) = \left\{
\begin{array}{lcr}
1 + \mbox{O}(y)\;,
&{\rm for}& y\ll 1 \\
\left( \frac{2}{\pi y} \right)^{1/2}+ 
\mbox{O}(y^{3/2})\;, &{\rm for}& y\gg 1
\end{array}
\right.
\label{B:flim}
\ee

As follows from (\ref{B:BDOSfin}), close to the boundary
the local density of states turns out to be
logarithmically enhanced as compared to the bulk one 
[c.f. (\ref{GF:DOS})]
\be
\bar{\rho}(y=0,s)
=\frac{2\pi\rho_0}{s\ln^2(1/s)}
\label{B:DOSfinbis}
\ee
(see also Appendix B).
Away from the boundary this enhancement is exhausted,
for the function $\bar{\rho}(y,s)$ decays as ($y \gg 1$):
\be
\bar{\rho}(y,s)
\simeq\frac{\sqrt{8\pi}\rho_0}{s\ln^2(1/s)}
\frac{1}{\sqrt{y}}
\label{B:DOSas}
\ee
It crosses over to the bulk value (\ref{GF:DOS}) at the
distances $y$ of the order of $\ln^2(1/s) \sim L_s$.
Thus the boundary influences the local density
of states on the scale of the order of the
correlation length.
A similar conclusion has been reached by AP
on the basis of the boundary
density of states distribution function \cite{AP}
(we refer to the case when the excitations
can not leave the sample).
In our case,
the difference with the standard localization problem is that
perturbations caused by the boundary decay into the bulk
on the scale $L_s$, not $\lambda_s$, and that they 
follow a power law, not an exponential decay.

\subsection{Density of states by renormalization group}

In the Introduction we have referred to 
real--space renormalization group (RG) calculations
for random spin models.
It is worth to investigate how this 
alternative approach works
for calculating the local density of states 
close to the boundary. 
We use the real space version of the RG,
introduced by Dasgupta and Ma\cite{Ma}
and extended by Fisher\cite{Fisher}, to cope with disordered
exchange constants in an Heisenberg chain. 
In fact, the random Heisenberg
model in the anisotropic XY limit and at zero magnetization 
is equivalent to a model of spinless fermions 
with random nearest-neighbor 
hopping integrals at half-filling. 
As we said, the random hopping has a 
particle-hole symmetry at half-filling, which prevents the 
generation of forward scattering processes. 
Therefore, this model is
exactly equivalent to the random backscattering model. 
Moreover,
it was shown \cite{Fisher} that the spin anisotropy is not a relevant
parameter, and that in the whole range $0\leq J_z \leq J_x=J_y$ 
the physical behavior is unchanged, including also the exponents
of the power--law decaying correlation functions. Hence, we
are going to consider the spin--isotropic case. 

The RG procedure consists in successive eliminations of pairs of
spins, which are coupled more strongly than the others.
The cut-off energy $\Omega$, 
which is rescaled downwards, is simply the largest exchange coupling
$\Omega=\mbox{Max}(J)$ at a given stage of the RG process. 
Once a pair of spins is decimated, 
an effective exchange is generated between the two
spins adjacent to the decimated bond. 
The probability $Q(y,\Gamma)$ that a bond of
length $y$ will be decimated at the scale 
$\Omega=\Omega_0 {\rm e}^{-\Gamma}$ 
has been calculated in Ref.\onlinecite{Fisher}
($\Omega_0$ being the cut-off energy
at the start of the RG process). 

Let us consider a semi--infinite chain. 
The probability $n(x,\Gamma)$ 
that the spin at site $x\in [0,\infty]$
is still free at the scale $\Omega$ 
satisfies the differential equation:
\begin{equation}
\frac{d \ln n(x,\Gamma)}{d\Gamma} = - \int_0^x dy \, Q(y,\Gamma)
-\int_0^\infty dy \, Q(y,\Gamma).
\label{dn}
\end{equation}
The two integrals on the right--hand--side imply
that the spin at $x$ gets bound to a spin at its left
or its right.
From the RG equations, one finds that $Q(y,\Gamma)$ at 
$\Gamma\gg 1$ is given by the inverse 
Laplace transform \cite{Fisher,FM2}
\begin{eqnarray}
Q(y,\Gamma) &=& \int_{c-i\infty}^{c+i\infty} \frac{dz}{2\pi i}
\frac{\sqrt{z}}{\sinh \left(\sqrt{z}\Gamma \right) }
{\rm e}^{zy/2}\nonumber \\
&=& \frac{\pi^2}{\Gamma^3} \sum_{m=1}^{\infty}
(-1)^{m+1} m^2 {\rm e}^{-\pi^2 m^2 y/(2\Gamma^2)}.
\label{Q}
\end{eqnarray}
From this equation one derives that
\[
\int_0^\infty dy Q(y,\Gamma) = \frac{1}{\Gamma},
\]   
and, for $1\ll x \ll \Gamma^2$, 
\[
\int_1^\Gamma d\Gamma' \int_0^x dy Q(y,\Gamma') \simeq 
\frac{1}{2} \ln\left( \frac{\pi^2 x}{2}\right).
\]
The limiting expressions of $n(x,\Gamma)$ can 
easily be obtained from the
above equations:
\[
n(x,\Gamma)\simeq
\left\{
\begin{array}{lcr}
\displaystyle
\frac{1}{\Gamma^2} &{\rm for} & x\gg \Gamma^2\\
\displaystyle
\frac{1}{\Gamma\sqrt{x}} &{\rm for} & 1\ll x \ll \Gamma^2\\
\displaystyle
\frac{1}{\Gamma} &{\rm for} & x\ll 1
\end{array}
\right.
\]
In the fermion language, $n(x,\Gamma)$ is proportional to the local density of
states $\rho(x,\epsilon)$ integrated up to the energy 
$\epsilon = \Omega_0 {\rm e}^{-\Gamma}$. We 
finally obtain 
\[
\rho(x,\epsilon)\sim\left\{
\begin{array}{lcr}
\displaystyle
\frac{1}{|\epsilon|\ln^3(1/|\epsilon|)} & {\rm for} & x\gg L_\epsilon\\
\displaystyle
\frac{1}{\sqrt{x}|\epsilon|\ln^2(1/|\epsilon|)} & {\rm for} & 1\ll x \ll 
L_\epsilon\\
\displaystyle 
\frac{1}{|\epsilon|\ln^2(1/|\epsilon|)} & {\rm for} & x\sim 1
\end{array}
\right.
\]
where the cross-over length $L_\epsilon = 2\ln^2(\epsilon)/\pi^2$. The 
resulting density of states is therefore 
in perfect agreement with the previous calculation.
An important observation is that from the RG analysis it
comes out quite naturally that the relevant correlation length 
in this problem is $L_\epsilon$. 
This is one of the main achievements of this approach, 
as we are going to discuss in
the following section. 
Before concluding this section, it is worthwhile
to briefly discuss the enhancement of the local density of states
close to a boundary in the light of the RG approach. 
The RG provides a quite simple picture of the ground state of the
random Heisenberg model, in terms of the so-called 
`random singlet state'\cite{Fisher}. 
With this we intend a state 
where each spin is coupled into a singlet with another spin, 
and the
longer the distance between the two coupled spins the lower the
excitation energy of the singlet. 
Hence, it is more or less obvious that a spin
close to a boundary has less 
probability to get coupled to another spin, 
which explains the enhancement of the 
density of states at low energy. 

\section{Discussion and Conclusions}
\label{sec:DC}

To summarize, in this paper we have investigated the 
single--particle Green function and the local
density of states in the presence of boundaries
for the random mass Dirac problem.
Let us discuss these findings in reverse.

For the density of states of the random
mass Dirac fermions is strongly energy
dependent (Dyson singularity),
it is not a surprise that the presence of the
boundary is essential. 
It is perhaps less obvious that the
density of states close to the boundary is strongly
enhanced as compared to the bulk one \cite{DCremark0}
\be
\bar{\rho}(x=0,s)/\rho(s)\simeq\frac{1}{4} \ln(1/s)\;.
\label{DC:DOS}
\ee

Regarding the spin systems, this result is relevant 
for the situation when one considers the
effect of especially strong impurities
(like lattice defects) in an already doped sample.
Indeed, the Dyson singularity itself readily
yields peculiar thermodynamic properties \cite{FM,NYG}.
For example, the linear magnetic susceptibility
diverges as $\chi(T) \sim 1/T\ln^2(1/T)$ \cite{DCremark1}.
Heeding (\ref{DC:DOS}),
one expects (larger) contributions of the order
of 
\[
\chi(T) \sim 1/T\ln(1/T)
\]
to the magnetic susceptibility due to strong impurities and, of course,
sample boundaries (the latter are especially important for powder
samples on which most of the experiments are presenty being conducted
\cite{LDop}).

According to the result (\ref{GF:Gfin}),
the single--particle Green function behaves as
\be
G^+(x,s)\sim \frac{\rho_0}{s\ln^2(-is)}\left(
\frac{1}{x}\right)^{3/2}\;,
\label{DC:G}
\ee
in the region $l\ll x \ll L_s$.
(In the region $x\gg L_s$ we expect the Green
function to cross over to a standard exponential
decay.)
It is worth noting that the
Green function (\ref{DC:G}) has a large
prefactor.
Therefore it is actually larger than the
bare Green function $G^{(0+)}\sim \rho_0$ on
the whole scale $l\ll x \ll L_s$.
The conclusion is that not only the density
of states is large (that is not surprising
for the zero--modes are packed around 
zero energy), but also the correlations are
strong. 
In fact, the one--particle correlations in the
disordered system are stronger that those in a
pure system (without the gap). 
This is only true below a certain 
energy scale, which vanishes with the
impurity concentration. 

The length scale $L_s \simeq l \ln^2 s$, below which
(\ref{DC:G}) is valid, can be inferred from the analysis of the
boundary effects. 
The same scaling relation of the length $x$ versus 
the energy $s$, 
$x \sim l \ln^2(s)$, was found in Ref.\onlinecite{George} by means of a 
replica method analysis of the Dirac fermion model, and also was derived by 
the RG investigation of the random Heisenberg model in Ref.\onlinecite{Fisher}. 
Surprisingly, it differs 
from the scaling relation that the
energy dependence of the localization length would suggest,
namely $x \sim \lambda(s) = l |\ln s|$. 
This is an evident
signature that the typical behavior differs from the average behavior,
the difference becoming large as $s\to 0$.
In other words, some physical quantities are dominated
by very rare events.

Notice that $L_s$ does not come out in 
any simple way from our asymptotic calculation of the Green function. 
The likely reason is that, once we
solve the recurrence relations for the Berezinskii blocks under the
restriction (\ref{GF:limit}), we are implicitly assuming that
$L_s\gg x$. This denies the access to the region $x\gg L_s$ where
the Green function decays exponentially. 
This issue ought to be further studied \cite{SFG}.
In fact, in the case when we knew the 
$x\gg L_s$ asymptotics, that is for 
the density of states in the presence of a boundary, we have been able to 
identify $L_s$.   

As we mentioned in the Introduction, also the RG method gives access to 
the correlation functions, although at equal times. In particular, the
equal-time spin-spin correlation function is known to decay at large
distances as \cite{Fisher}
\[
\langle S^i(x)S^i(0) \rangle \sim \frac{1}{x^2},
\]
($i=x,y,z$). One can show\cite{FM2} that also the equal-time Green function 
in the equivalent random hopping tight-binding model decays as $1/x^2$.
This result is, in fact, compatible with (\ref{DC:G}). Indeed, (\ref{DC:G})
is only valid for $x\ll L_s$, or, equivalently, for
$s\ll s(x)=\exp(-\sqrt{x/l})$. Therefore, neglecting exponentially decaying 
terms,
we find
\begin{equation}
G^+(t=0,x) \sim \int_0^{s(x)} ds G^+(x,s) \sim \frac{\rho_0}{x^{3/2}}
\int_0^{s(x)} \frac{ds}{s\ln^2s} \sim \frac{1}{x^2},
\label{DC:Gt}
\end{equation}
in agreement with the RG results. 
Hence our results not only reproduce the
equal-time behavior of the correlation functions found by RG, but also
allow us to determine their energy dependence.  

According to GM \cite{AAG}, the two--particle
correlations follow the same pattern as (\ref{DC:G}). 
The fact that the disorder actually enhances 
correlations (at low energy scale) throws the
light onto the experimental findings of
an antiferromagneting ordering in the spin--Peierls
compounds building up upon doping \cite{SP}.
Indeed, the system without the spin--phonon
interactions is known to be unstable with respect to
the antiferromagnetic ordering.
In a pure system, however, the latter is prevented
by the spin gap. 
The doping effectively creates states in the gap and
induces correlations that are even stronger
than those in a pure system without the spin gap.
Hence the antiferromagnetism is promoted by doping.
This matter, though, is to be investigated in more detail,
especially with respect to the energy scales involved 
(we expect the relevant scale be associated with
the fluctuations of the 
density--density correlator) \cite{SFG,DCremark2}.

To conclude, in this paper
we have analyzed single--particle
properties of the random mass Dirac problem
in some detail.
The lack of knowledge of the distribution
functions (for the density of states, etc.),
however, calls upon further exploration of the problem.

\section{Acknowledgements}

We are thankful to Y. Chen, S. Gogolina, and R. M\`elin for 
interesting and helpful discussions.
M.F. was supported by INFM, under project HTCS.
A.O.G. was supported by the EPSRC of the United Kingdom.
M.S. was supported by the Gottlib Daimler- 
und Karl Benz- Stiftung.

\section{Appendix A}

In order to evaluate the integrals in (\ref{GF:Gspace},
\ref{GF:Qspace}), which read
\be
F(y)=\int\limits_0^\infty dz
K_0(z)\int\limits_0^\infty d\tau
\tau\tanh \left( \frac{\pi\tau}{2} \right) 
K_{i\tau}(z) e^{-\frac{\tau^2}{2} y}\;,
\label{AA:F}
\ee
we make use of Nicholson's integral representation of the
product of MacDonald functions \cite{W}
\be
K_0(z)K_{i\tau}(z)=2\int\limits_0^\infty dt
\cos (\tau t)K_{i\tau}(2z \cosh t)\;.
\label{AA:Nic}
\ee
Upon substituting (\ref{AA:Nic}) into (\ref{AA:F})
one observes that the $z$--integral and the $t$--integral,
if taken in succession, are the table ones \cite{GR}.
Hence (\ref{GF:F}).

\section{Appendix B}

Exactly at the sample boundary the local density of states 
can be explicitly evaluated. 
Indeed, owing to the boundary condition (\ref{B:bc}), 
the formula (\ref{B:DOSbloks}) for the density of states 
simplifies so that the summation over the number of lines 
is easily performed [c.f. (\ref{GF:Rm})]:
\[
\bar{\rho}(x=0,\epsilon)/\rho_0=1+ 2{\rm Re}
\sum\limits_{m=1}^\infty R_m=
-1+2{\rm Re}\left\{
\frac{2{\rm e}^{is/4}}{K_0(-is/4)}\int\limits_0^\infty dt
\sqrt{\frac{t+1}{t}}{\rm e}^{ist/2}\right\}\;.
\]
The above integral is a table one \cite{GR}, 
so that we finally obtain
\be
\bar{\rho}(x=0,\epsilon)/\rho_0={\rm Re}\left\{
\frac{K_1(-is/4)}{K_0(-is/4)}\right\}
\label{AppDOSxnot}
\ee
[the $s\to 0$ limit of this formula is in agreement with
Eq.(\ref{B:BDOSfin})].

\newpage

\newpage
\begin{figure}
\caption{The elementary vertices of the Berezinskii diagram technique
contributing the factors (a) $-1/(2l)$, (b) $-1/l$, 
(c) $-(1/l)\exp\left[-i(\epsilon/v)x\right]$,
(d) $-(1/l)\exp\left[+i(\epsilon/v)x\right]$.}
\label{berfigver}
\end{figure}

\begin{figure}
\caption{The configurations of the electron line end points for the
single--particle Green function.}
\label{gffigend}
\end{figure}

\begin{figure}
\caption{This figure illustrates the definition of Berezinskii blocks for the 
diagrams
with end point configuration shown in Fig.
\protect\ref{gffigend}(a).}
\label{gffigbl}
\end{figure}

\begin{figure}
\caption{The process of joining the elementary vertices for deriving the 
recurrence relation for the right--hand--side blocks $\tilde{R}_m(x)$.}
\label{gffigpr}
\end{figure}

\begin{figure}
\caption{Two fourth order mirror imaged (coinciding end points)
diagrams contributing to the Green function.}
\label{gffigmir}
\end{figure}

\begin{figure}
\caption{The plot of the function $F(y)$ determining the spatial decay of the
single--particle Green function.}
\label{gffigF}
\end{figure}

\begin{figure}
\caption{The new element of the diagram technique in the presence of the 
boundary --
the boundary vertex.}
\label{bfigbv}
\end{figure}

\begin{figure}
\caption{An example of a diagram involving the boundary scattering.}
\label{bfigex}
\end{figure}

\begin{figure}
\caption{The plot of the function f(y) determining the spatial decay of the 
local
density of states away from the boundary.}
\label{bfigf}
\end{figure}


\begin{references}
\bibitem{Mremark1}
More precisely, the Dirac model occurs in the total spin
spin sector, while there also are two massive Majorana fermions
in the `spin--flavor' sector: see discussion in 
Ref.\protect\onlinecite{SNT}.
\bibitem{Mremark2}
A more formal way to see this is to track
the spatial dependence of the average charge
phase field (which effectively takes care of the site
numbering) in the definition of the staggered magnetization 
$\vec{n}(x)$ \protect\cite{NYG}.
\bibitem{Berremark1}
This does not lead to `over--counting'.
Indeed, after averaging over the disorder, one finds
contributions to the Green function given as 
unlimited integrals over the spatial coordinates.
Each of these contributions splits into a
sum of terms with the spatial integration taken 
over a given $x$-ordered domain. These correspond
to different arrangements of the `interaction' vertices 
so that each of them is represented by an unique diagram.
\bibitem{DCremark2}
Theoretical
outlook for the spin--ladder systems,
where an antiferromagnetic tendency upon
doping has recently been discovered 
\protect\cite{LDop},
is less clear. 
The reason is that in these systems, unlike the spin--Peierls
systems 
\protect\cite{FM}, 
the staggered magnetization
is not directly related to the density--density
correlation function of the Dirac fermions,
but it is a more complex object 
\protect\cite{SNT}.
\end{references}
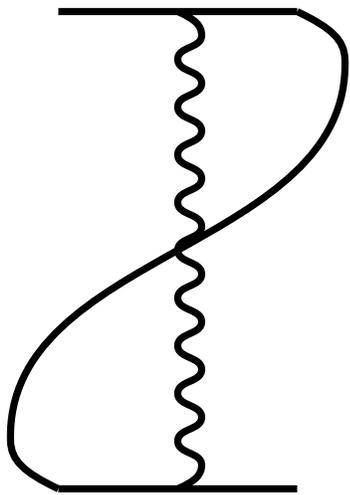
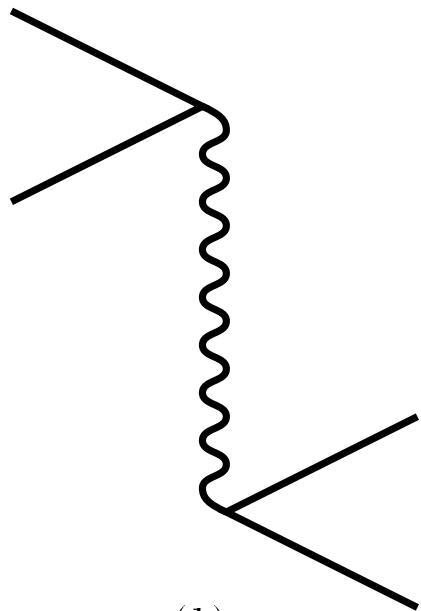
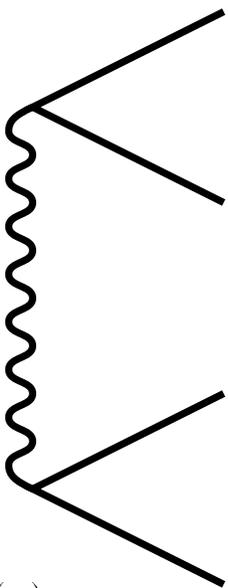
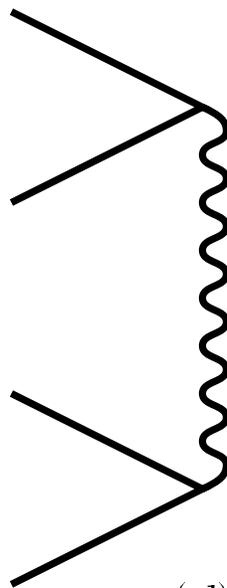
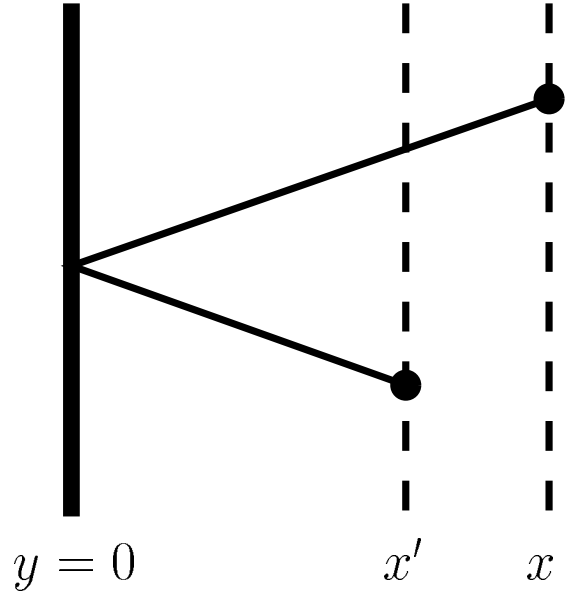
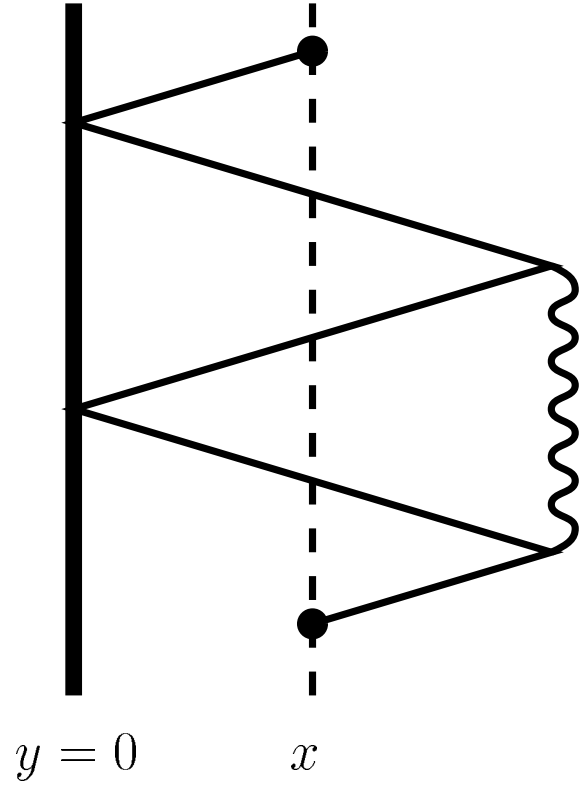
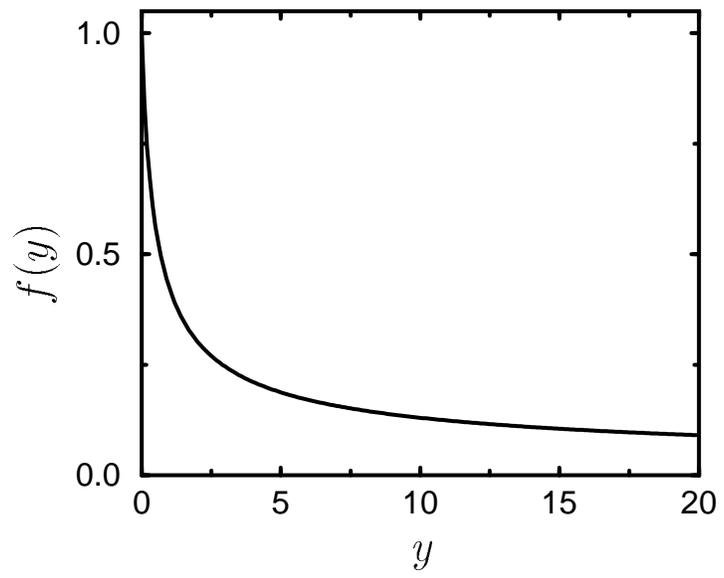
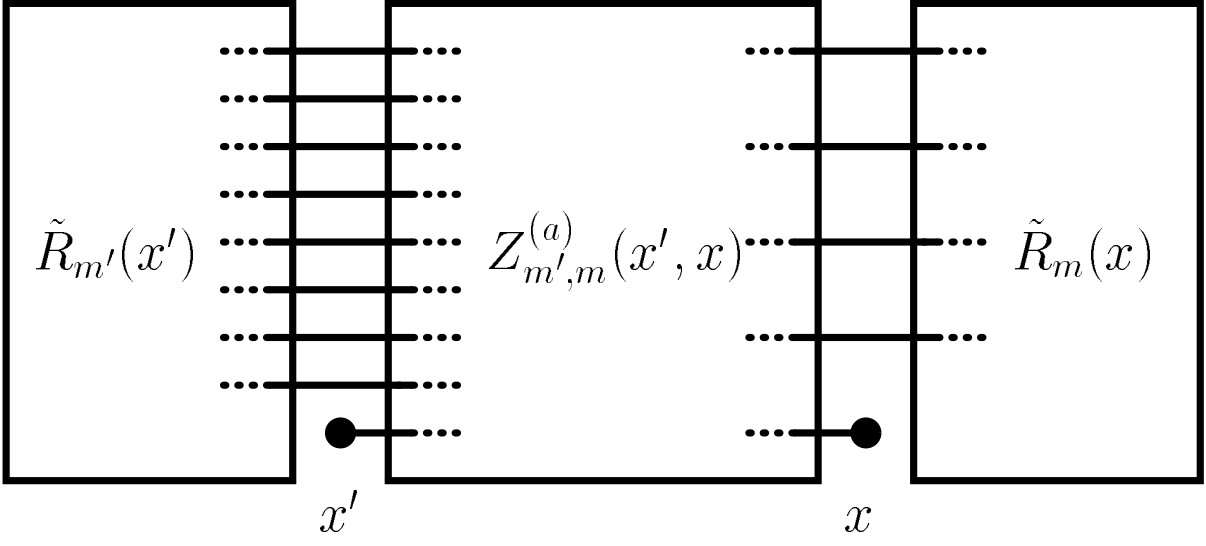
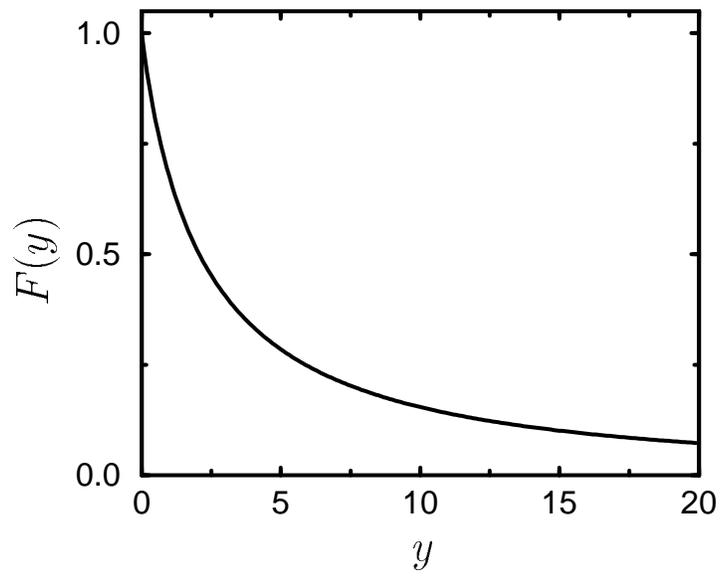
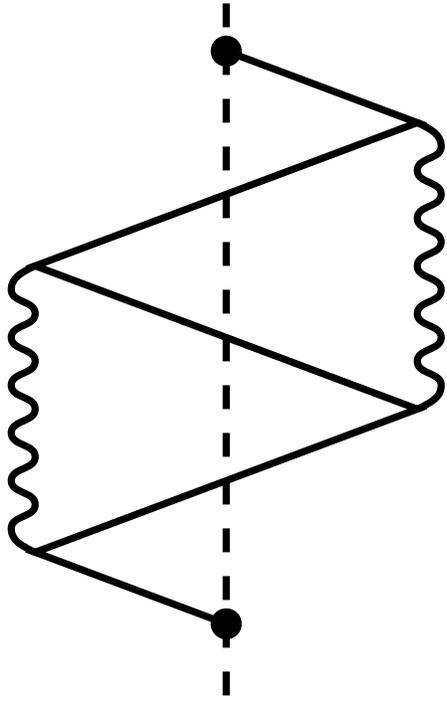
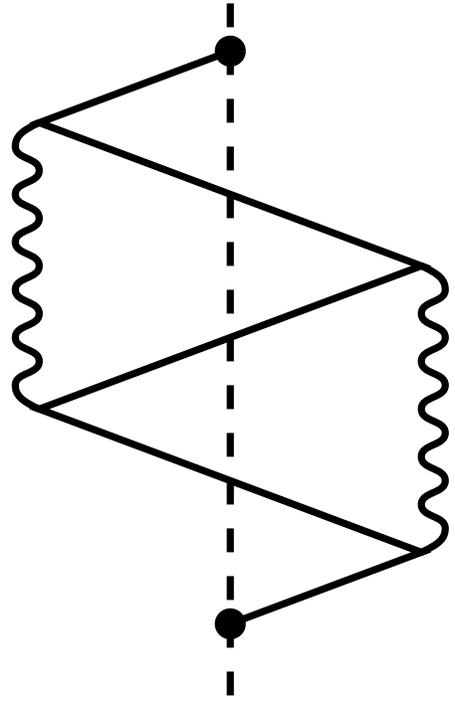
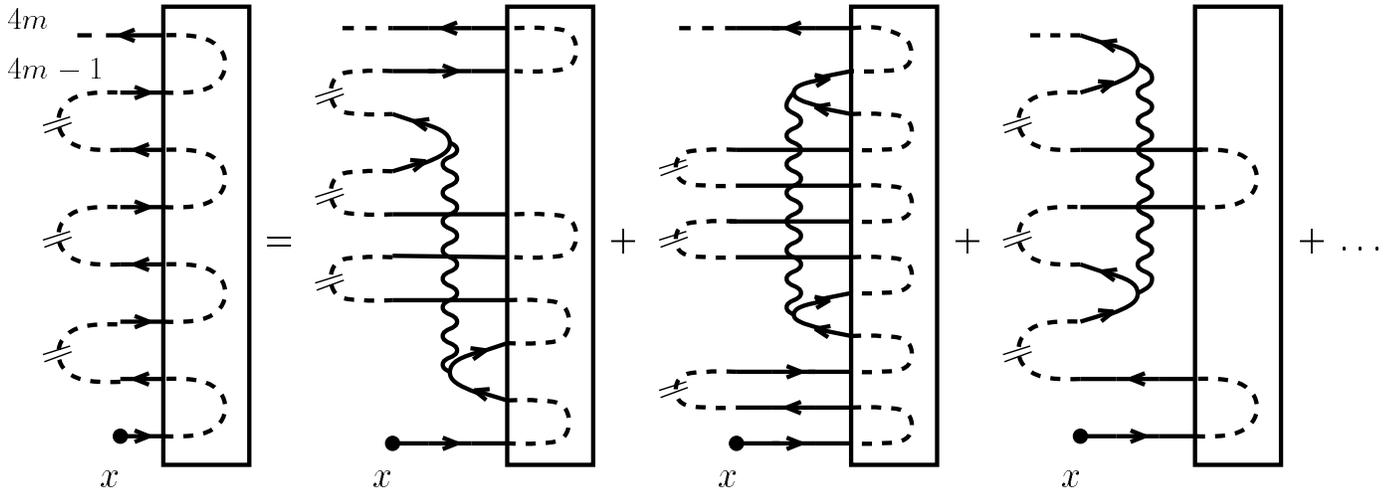
\end{document}